\documentstyle[emulateapj]{article}

\def\eq#1{equation~(\ref{#1})}
\def\Eq#1{Eq.~\ref{#1}}
                                   
\def\keywords{}

\def\aap{A\&A}
\def\apjl{ApJ}
\def\apjs{ApJS}

\begin{document}

\def\newpage{\vfill\eject}
\def\vs{\vskip 0.2truein}
\def\msun{M_\odot}
\def\rsun{R_\odot}
\def\met{[M/H]}
\def\vi{(V-I)}
\def\mtot{M_{\rm tot}}
\def\mhalo{M_{\rm halo}}
\def\pp{\parshape 2 0.0truecm 16.25truecm 2truecm 14.25truecm}
\def\la{\mathrel{\mathpalette\fun <}}
\def\ga{\mathrel{\mathpalette\fun >}}
\def\fun#1#2{\lower3.6pt\vbox{\baselineskip0pt\lineskip.9pt
  \ialign{$\mathsurround=0pt#1\hfil##\hfil$\crcr#2\crcr\sim\crcr}}}
\def\kpc{{\rm kpc}}
 \def\Mpc{{\rm Mpc}}
\def\kmsec{{\rm km/sec}}
\def\ibl{{\cal I}(b,l)}
\def\kms{{\rm km}\,{\rm s}^{-1}}
\def\mjup{M_{\rm J}}
\def\rjup{R_{\rm J}}
\def\rp{R_{\rm p}}
\def\rs{R_*}
\def\au{\rm AU}
\def\tt{t_{\rm T}}
\def\ci{\cos{i}}
\def\cimin{\ci_{\rm min}}
\def\apsf{\Omega_{\rm PSF}}
\def\pt{P_{\rm T}}
\def\pw{P_{\rm W}}
\def\pstn{P_{\rm S/N}}
\def\ptot{P_{\rm tot}}
\def\tw{T_{\rm W}}
\def\np{n_{\rm P}}
\def\qmin{Q_{\rm min}}
\def\ndet{N_{\rm det}}

\lefthead{GAUDI}
\righthead{PLANETARY TRANSITS TOWARD THE GALACTIC BULGE}
\title{Planetary Transits Toward the Galactic Bulge}
\author{B. Scott Gaudi}
\affil{Department of Astronomy, The Ohio State University,
Columbus, OH 43210, USA}
\authoremail{gaudi@astronomy.ohio-state.edu}


\begin{abstract}
The primary difficulty with using transits to discover extrasolar planets is the
low probability a planet has of transiting its parent star.  One
way of overcoming this difficulty is to search for transits in dense
stellar fields, such as the Galactic bulge.  Here I estimate the
number of planets that might be detected from a monitoring campaign
toward the bulge.  A campaign lasting 10 nights on a 10
meter telescope (assuming 8 hours of observations per night and a
5'x5' field of view) would detect about
$100$ planets with radius $\rp=1.5~\rjup$, or about $30$ planets with
$\rp=1.0~\rjup$, if the
frequency and distribution of planets in the bulge is similar to that
in the solar neighborhood.  Most of these planets will be discovered around
stars just below the turn-off, i.e. slightly evolved
G-dwarfs.  Campaigns involving 1- or 4-m class telescopes are unlikely
to discover any planets, unless there exists a substantial population
of companions with $\rp > 1.5~\rjup$.  

\end{abstract}


\setcounter{footnote}{0}
\renewcommand{\thefootnote}{\arabic{footnote}}

\section{Introduction}

The search for extrasolar planets has garnered enormous attention in
recent years, due primarily to the successful implementation of radial
velocity searches (\cite{mandq1995}, \cite{mandb1996}).  These searches have led to the
discovery of a population of massive, close-in planets with orbital
separations of $a \la 0.1~\au$.  Recently, it was discovered that one
such planet, the companion to HD 209458, also transits its parent star 
(\cite{charbon2000}; \cite{henry2000}), yielding a measurement of the
mass, radius, and density of the companion.   

Clearly, transit observations can be used to extract additional information
about known companions.  The {\it discovery} of an extrasolar planet using transits,
however, has remained elusive.  There are two primary difficulties
with detecting planets with transits.  First, the photometric
requirements are quite stringent: a planet of radius $\rp\le\rjup$
(where $\rjup$ is the radius of Jupiter)
transiting an primary of radius $\rs=\rsun$ would produce a fractional
deviation of $\la 1\%$ during the course of the transit.  Second, the
probability that a planet will transit its parent is small: 
for a planet with separation $\ge 0.05~\au$ orbiting a
star with $\rs=\rsun$, the probability is $\la 10\%$.  Several methods for dealing will
the small probability have been proposed.  For instance, one can monitor eclipsing
binary stars, where the orbital plane is known to be (nearly) perpendicular to
the sky (\cite{deeg}).  

Another way of overcoming this small probability is to simply monitor
many stars simultaneously.  This can be done by employing a camera with
a large field-of-view, or by monitoring very dense stellar fields.
Here I focus on the latter possibility.  Specifically I determine the number of
planets that might be detected in a campaign monitoring stars toward
the Galactic bulge. 

\section{Formalism}

The flux of a star being occulted by a planet is given by,
\begin{equation}
F(t)=F_0 [\delta(t)+1] + F_b,
\label{eqn:flux}
\end{equation}
where $F_0$ is the unocculted flux of the star, $F_b$ is the total
flux from any unrelated sources, and $\delta(t)$ is the fractional
deviation of the flux due to the transit, which depends on the radius
of the planet relative to the star, the inclination angle, $i$, and the
limb-darkening of the star (\cite{sackett1999}).  For a small planet
($\rp \ll \rs$) and no limb-darkening, $\delta=(\rp/\rs)^2\Theta(1-\tau)$, where $\Theta(x)$ is
the step function, and $\tau$ is a normalized time, 
$\tau \equiv {({ t-t_0}) / \tt}$.
Here $t_0$ is the time of the midpoint of the transit, and $\tt$ is
one-half the transit duration, which for circular orbits is,
\begin{equation}
\tt= {P \over 2\pi}{\rm arcsin}\left(\sqrt{  \left({{\rs+\rp}\over
a}\right)^2 - \cos^2{i}}\right).
\label{eqn:tt}
\end{equation}
In reality, $\delta$ depends very sensitively on $\rp$ and $\ci$, and
less so on the limb-darkening.  I will therefore use the explicit
form for $\delta$ given in Sackett (1999), but assume no limb-darkening.  

Since the proposed search for planets will be carried out in dense
stellar fields, and transits produce time-dependent variations in
the flux of the stars, the data will likely be reduced with
image-subtraction techniques (\cite{tandc1996}, \cite{andlup1998}).
With image-subtraction, one measures only the time variable
portion of the flux, 
${\tilde F(t)}=F_0[\delta(t)]$.

\section{Detection Probability}

There are three requirements to detect a planet
of separation $a$ and radius $\rp$ around a star of mass $M$, radius
$\rs$ and flux $F_0$.  These are: (1) the planet must
transit the star, (2) at least two transits must occur during the
time when observations are made, and (3) the
transit must cause a detectable deviation in the light curve.  If the
duration of the transit is much smaller than the window of
observations, than these three requirements can be considered
independent.  

For a planet to transit its parent star, it must have an inclination
angle $\ci \le \cimin \equiv {(\rs+\rp)/a}$.
The probability $\pt$ that a planet will transit its parent star is therefore,
\begin{equation}
\pt = {{\int_0^{\cimin} {\rm d}(\ci)}\over{\int_0^1
{\rm d}(\ci)}}={{\rs+\rp}\over a}.
\label{eqn:pt}
\end{equation}

Consider a campaign lasting $N$ nights with $\tw$ hours per night.
Defining $t=0$ as the beginning of the first night, then the times
when observations are possible on (integer) night $n$ satisfy ${\cal
T}(t)=|t - n\lambda - \tw/2| - \tw/2 \ge 0$, where $\lambda=1~{\rm day}$.
Both the first transit occurring at time $t_1$ and the second
transit at time $t_2$ must satisfy this relation on some (integer) nights $n_1$ and
$n_2$.  Note $n_1 \ge n_2$.  The time $t_2$ is given by $t_2=t_1+\np
P$, where $\np$ is the number of periods between $t_1$ and $t_2$.
Since $t_1$ can occur anywhere in the time span $0\le t_1 \le P$, then
the probability that both transits will occur during the window(s) of
observations is,
\begin{equation}
\pw = {1 \over P} \int_0^P {\rm d}t_1 
\Theta[{\cal T}(t_1)] 
\Theta[{\cal T}(t_1+\np P)],
\label{eqn:pw}
\end{equation}
for any combination of $n_1=0,1,...,N-1$, $n_2=n_1,n_1+1,....,N-1$, and $\np=1,2,...,N\lambda/P$.

Finally, consider a transit of duration $2\tt$ that occurs well within the observing
window.  Assuming that the transit is monitored continuously with a
telescope that records $\alpha$ electrons per second per unit flux, the
total signal-to-noise of the transit is,
\begin{equation}
Q= ( 2 \alpha \tt)^{1/2} { F_0 \over [S_{\rm tot}\Omega_{\rm
{PSF}}+F_0]^{1/2}} {\left( \rp \over \rs \right)^{2}} G,
\label{eqn:stn}
\end{equation} 
where $S_{\rm tot}=S_{\rm sky}+S_{\rm back}$ is the total surface
brightness (sky + unresolved background), $\Omega_{\rm {PSF}}$ is the
area of the PSF, and the function $G$ is defined as,
\begin{equation}
G \equiv \left({\rs \over \rp}\right)^2 
\left[ {1 \over 2}  \int_{-1}^{1} {\rm d}\tau [\delta(\tau)]^2 \right]^{1/2},
\label{eqn:gsquared}
\end{equation}
and depends on $\rp/\rs$, $\ci$, and the limb darkening of the star.
For $\rp \ll \rs$ and no limb-darkening, $G=1$.  Note that an implicit
assumption in \eq{eqn:stn} is that $\delta \ll 1$.  For a transit to
be detectable, $Q$ must exceed some minimum threshold, $\qmin$. 
Integration over $\ci$ then
defines the probability that a transit will satisfy the signal-to-noise
requirement,
\begin{equation}
\pstn = (\cimin)^{-1} \int_0^{\cimin} {\rm d}(\ci) \Theta(Q-\qmin).
\label{eqn:pstn}
\end{equation}
The total detection probability is then just 
$\ptot=\pt \times \pw \times \pstn$.

Consider a population of stars with luminosity function
$\Phi (F_0)$ (in units of number per area), mass-flux relation
$M(F_0)$ and radius-flux relation $R(F_0)$.  Assuming a fraction $f$
of these stars have planets of radius $\rp$ distributed according to
${\cal F}(a){\rm d}a$ (which I will assume is independent of $F_0$), then the number of
planets detected in a field of view of area $\Omega_{\rm CCD}$ is,
\begin{equation}
\ndet = f \Omega_{\rm CCD} \int {\rm d} a {\cal F}(a) 
\int {\rm d}F_0 \ptot \Phi(F_0).
\label{eqn:ndet}
\end{equation}

\section{Application to the Galactic Bulge}

Before applying the results of \S\ 2 to the Galactic bulge, 
several assumptions must be made about the population being
monitored, and also the telescope and observational setup. I will
consider observations in the $I$-band, which provides a good
compromise between dust extinction and high sky background.  I construct
an $I$-band luminosity function by combining the determination toward Baade's window
by Holtzman et~al.\ (1998) on the bright ($M_I\le 9$) end with the local
M-dwarf luminosity function as determined by Gould, Bahcall \& Flynn
(1997) for the faint end.  I normalize the latter to agree with
Holtzman et~al.\ (1998) at $M_I = 7.25$.  I adopt a distance modulus
of $14.52$ and an extinction of $A_I=1.0$, appropriate for Baade's
window.  For $M(F_0)$
and $R(F_0)$, I use the 10~Gyr, solar metallicity isochrone of
Girardi et~al.\ (2000).  These relations are shown in Figure 1.
Varying the age and/or metallicity of the population within
a reasonable range does not affect the results substantially (see
Fig.\ 1). 

\centerline{{\vbox{\epsfxsize=9.0cm\epsfbox{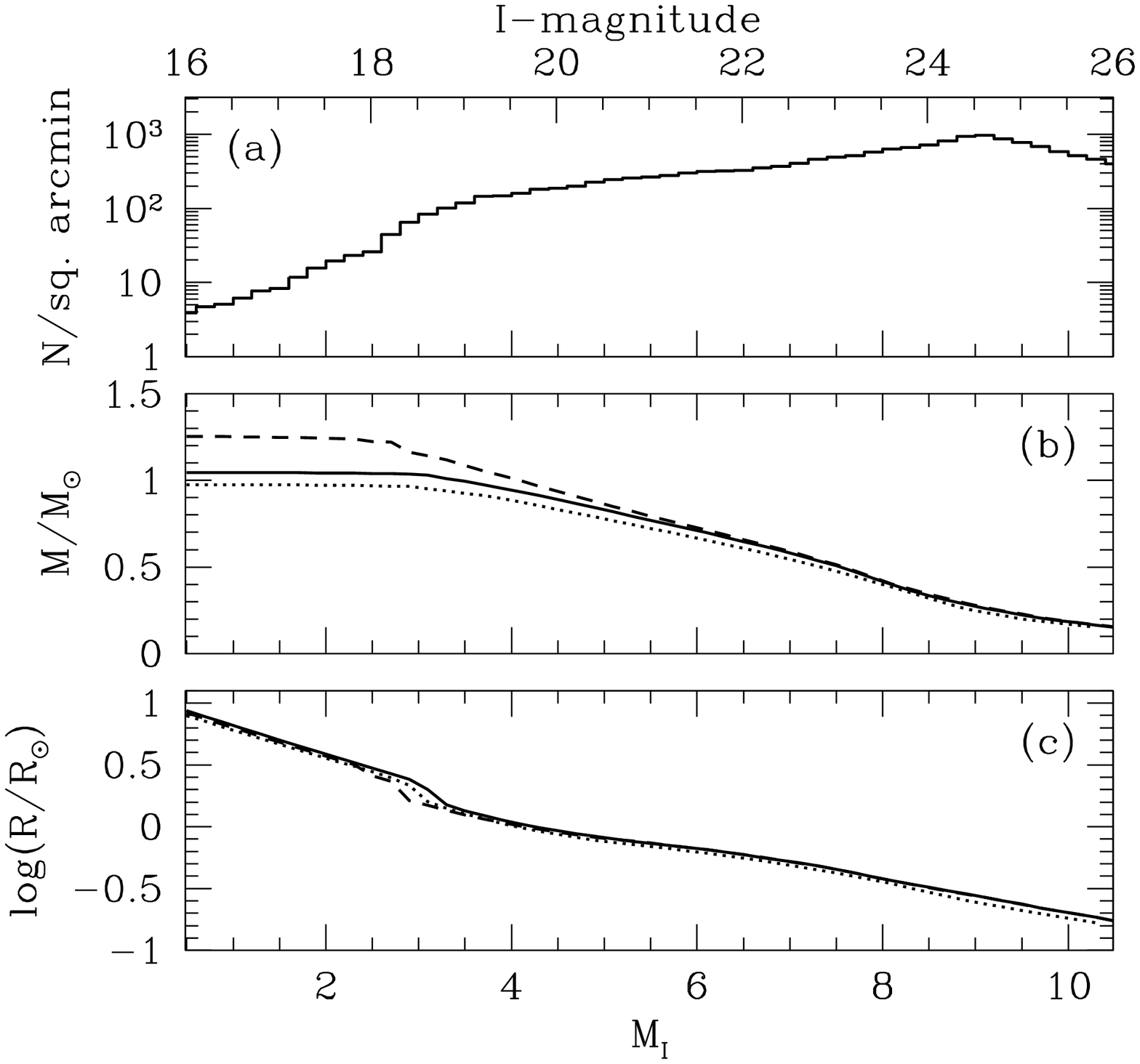}}}}
{\footnotesize {\bf FIG. 1}
(a) The adopted luminosity function (LF) for the Galactic bulge,
constructed from the Holtzman et~al.\ (1997) LF for $M_I\le 9$, and
the Gould et~al.\ (1997) LF for $M_I \ge 9$.  (b) The solid line shows
the adopted mass-luminosity
relation, as determined from the 10 Gyr, solar-metallicity isochrones
of Giraldi et~al.\ (2000).  Also shown are the mass-luminosity
relations for a 5 Gyr and solar-metalliticity population (dashed line)
and a 10 Gyr
and half-solar metallicity population (dotted line). (c) The radius-luminosity relation, as
determined from the same isochrones as in panel (b).  The bottom axis
shows $M_I$, while the top axis shows $I$, assuming a distance modulus
of 14.52 and $A_I=1.0$.
}
\bigskip

I assume that $S_{\rm sky}=19.5~{\rm mag~arcsec^{-2}}$, $\Omega_{\rm
PSF}=\pi \theta^2$, where $\theta$ is the seeing, and 
$\alpha=600(D/10{\rm m})^2~e^{-}~s^{-1}$ at $I=20$, 
where $D$ is the telescope diameter.  In the crowded fields toward the
Galactic bulge, the surface brightness $S_{\rm back}$
due to unresolved sources will depend strongly on the seeing.  To 
estimate $S_{\rm back}$, I first use the LF to determine the magnitude
at which the sources become unresolved, i.e., I determine the 
$I_{\rm min}$ such that all source brighter than $I_{\rm min}$
contribute on average one star per seeing disk.  
Then $S_{\rm back}$ is just the total
surface brightness due to all stars fainter than $I_{\rm min}$. 

These assumptions can now be used, along with the results of \S\ 2 to
determine the number of planets that may be detected in a monitoring
campaign toward the Galactic bulge as a function of the radius and
separation of the planet, and to explore the effects of the diameter $D$
of the telescope, the seeing, $\theta$, and the total number of nights
$N$ the field is monitored,  on the number of detections.

\centerline{{\vbox{\epsfxsize=9.0cm\epsfbox{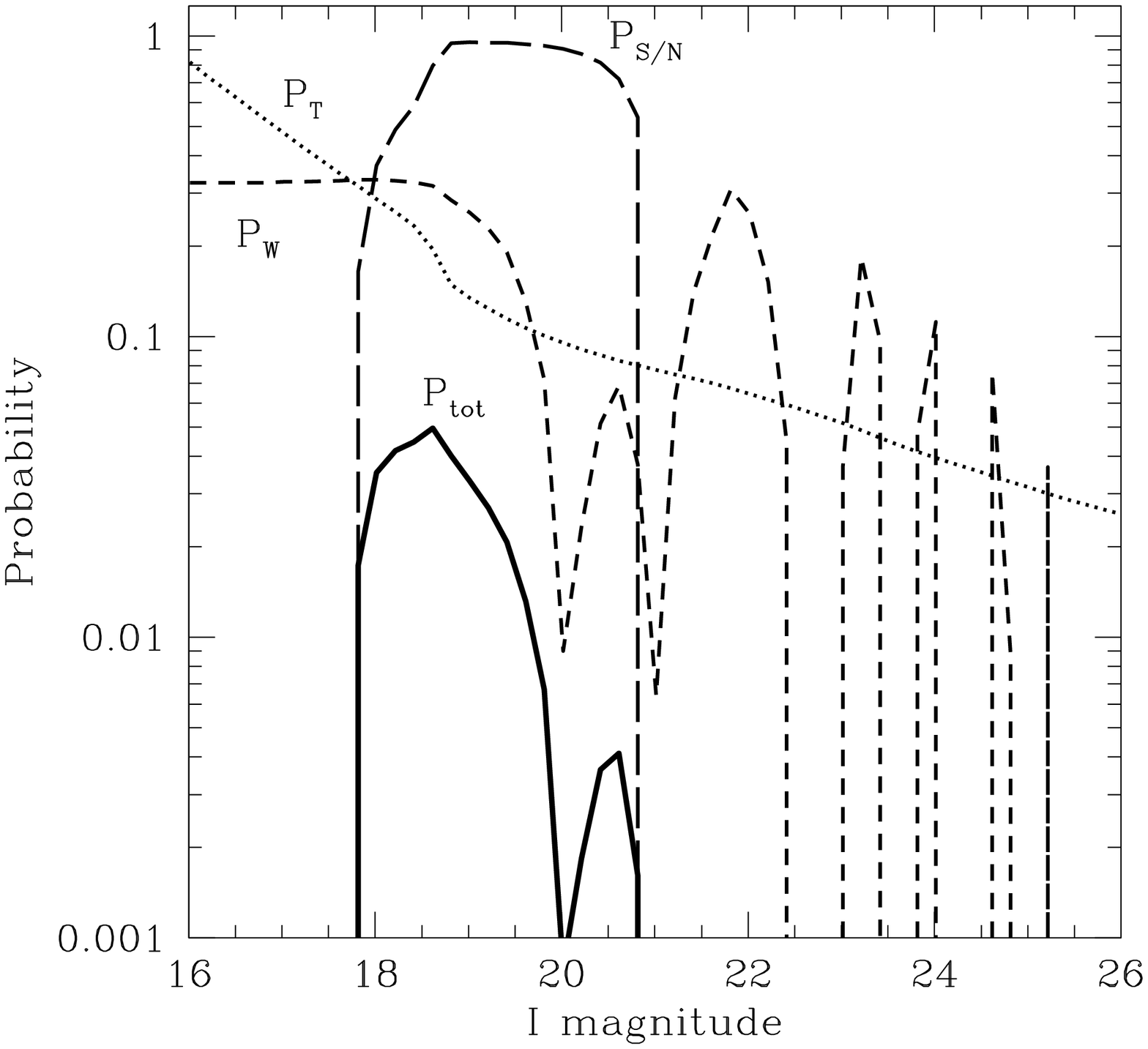}}}}
{\footnotesize {\bf FIG. 2}
The detection probability as a function of $I$ magnitude for a planet
with radius $\rp=1.0~\rjup$ and separation $a=0.05~\au$, or period
$P=4.08[M(I)/\msun]^{-1/2}~{\rm days}$,
assuming 10 nights of 8 hours per night on a 10m telescope with
0.75'' seeing.   The solid
curve shows the total detection probability, which is the given by 
$\ptot=\pt\times\pw\times\pstn$, where $\pt$ is the probability that
the planet will transit its parent star, $\pw$ is the probability that
two transits will occur during observation windows, and $\pstn$ is the
probability that the transit will have total signal-to-noise $>10$.  
}
\bigskip

Figure 2 shows the total detection probability $\ptot$ for a planet of radius
$\rp=1~\rjup$ and separation $a=0.05~\au$ as a function of $I$ magnitude, under the
assumptions of 10 nights of 8 hours per night on a 10m telescope with
$\theta=0.75''$, and a minimum signal-to-noise of $\qmin=10$
for a detection.   For planets orbiting stars slightly fainter than the
turn-off, $19 \le I \le 21$, almost all transits occurring during the
windows of observation create significant ($Q\ge \qmin =10$) transits,
i.e. $\pstn\sim 1$.   For $I\le19$, the radii of the sources rapidly
increases, rendering the transits undetectable.  For $I\ge 21$, the
sources produce too few photons to pass the signal-to-noise criterion.
For this particular separation, $a=0.05~\au$, the probability  $\pw$ that the
planet will transit twice during the windows of observation drops precipitously
for $20 \le I \le 21$, since the period of the planet,
$P=4.08[M(I)/\msun]^{-1/2}~{\rm days}$, moves into ``anti-resonance'' with the observation
window, $\tw=8~{\rm hours}$.  However, such effects will approximately average out
when a range of separations is considered.

The number of planets detected during a monitoring campaign can now be
determined by integrating over the luminosity function of the sources
and the separation of the companions (c.f.\ \Eq{eqn:ndet}). This 
requires knowledge of the frequency and distribution of planetary
companions to the bulge sources.
Obviously, little is known about planetary companions to bulge stars.
However, radial velocity surveys do provide information on the  
frequency and distribution of planetary companions to solar-type stars in the local
neighborhood.  Cumming, Marcy \& Butler (1999) performed a statistical
study of 74 solar-type stars from the Lick radial velocity
survey.  Of these, 2 had confirmed planetary ($M_{\rm p} \le 10~\mjup$)
companions with separations $\le 0.1~\au$, i.e. $3\pm 2\%$ of the
sample.  Furthermore, they note that the distribution in orbital
radius shows a ``piling-up'' toward small orbital radii, but that
this trend is not statistically significant.  It does hint, however, that
the distribution in $a$ may not be uniform.  
I will therefore assume that $f=1\%$ of all stars have planetary
companions distributed uniformly in $\log{a}$ between $0.01$ and $0.1~\au$.  While
this may not accurately reflect the frequency and distribution of
planets in either the solar neighborhood or the bulge, it is at least
consistent with available observations.

\centerline{{\vbox{\epsfxsize=9.0cm\epsfbox{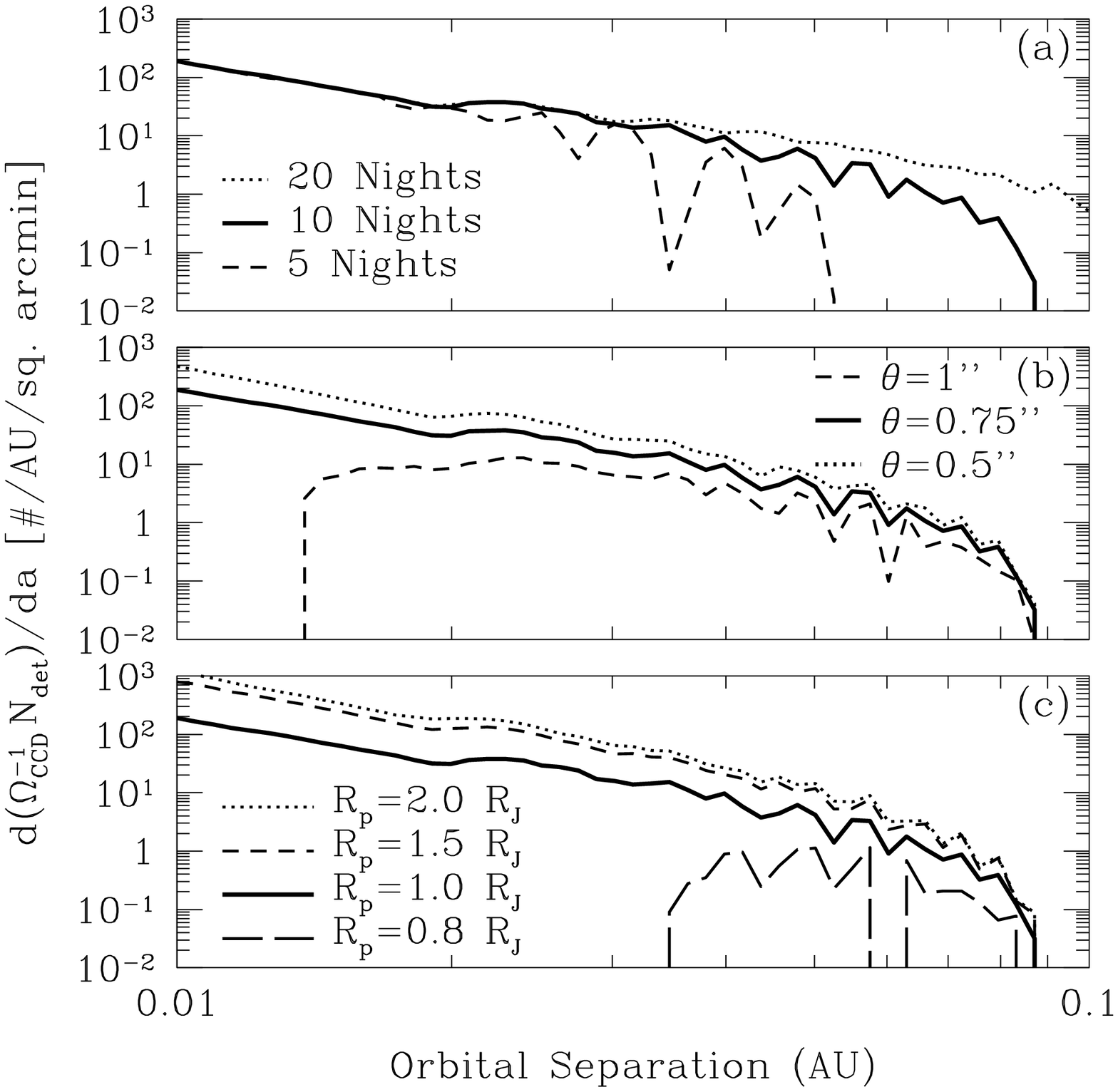}}}}
{\footnotesize {\bf FIG. 3}
The number of planets detected per AU per square arcminute monitored as a
function of orbital separation, assuming that 1\% of all stars have planets
distributed uniformly in $\log{a}$ between $a=0.01$ and $0.1$.
  All parameters are held constant at 
$\rp=1.0~\rjup$, $a=0.05~\au$, $N=10~{\rm nights}$, $\tw=8~{\rm
hours}$, $D=10{\rm m}$, and $\theta=0.75''$, unless otherwise stated.  
(a) The effect of varying the
total number of nights.  (b) The effect of varying the seeing,
$\theta$.  (c) The effect of varying the radius of the planet, $\rp$.
}
\bigskip

Figure 3 shows the differential distribution of the number of
detected planets per unit area, ${\rm d}(\ndet \Omega_{CCD}^{-1})/{\rm
d}a$, as a function of $a$ for $\rp=1~\rjup$, assuming $f=1\%$,
${\cal F}(a) \propto 1/a$, and the fiducial campaign with parameters
$N=10$, $\tw=8$~hours, $D=10$m, and $\theta=0.75''$.  Each panel 
shows the result of varying $N$, $\theta$, and $\rp$.  Decreasing the
duration of the campaign to $N=5$~nights will not substantially
decrease the number of detections: most of the planets lost will be at
large orbital separations, where the detection efficiency and frequency
are already low.  Similarly, doubling the number of nights will not
substantially enhance the number of detections, although it enables
the detection of planets at orbital separations larger than $0.1~\au$.
The number of detected planets depends quite crucially on the seeing:
increasing the seeing increases the number of unresolved sources, and
therefore adds to the background flux.  As $\theta$ increases, the
signal-to-noise degrades, and transits quickly fall below the minimum
detectable threshold.  Thus detections are lost, and preferentially so
for smaller separations (where the duration of the transits are
shorter).  Conversely, improving the seeing dramatically increases the
number of detections.  Therefore, transit searches toward the Galactic bulge should be carried
out at good sites with seeing better than 1''.
  
\centerline{{\vbox{\epsfxsize=9.0cm\epsfbox{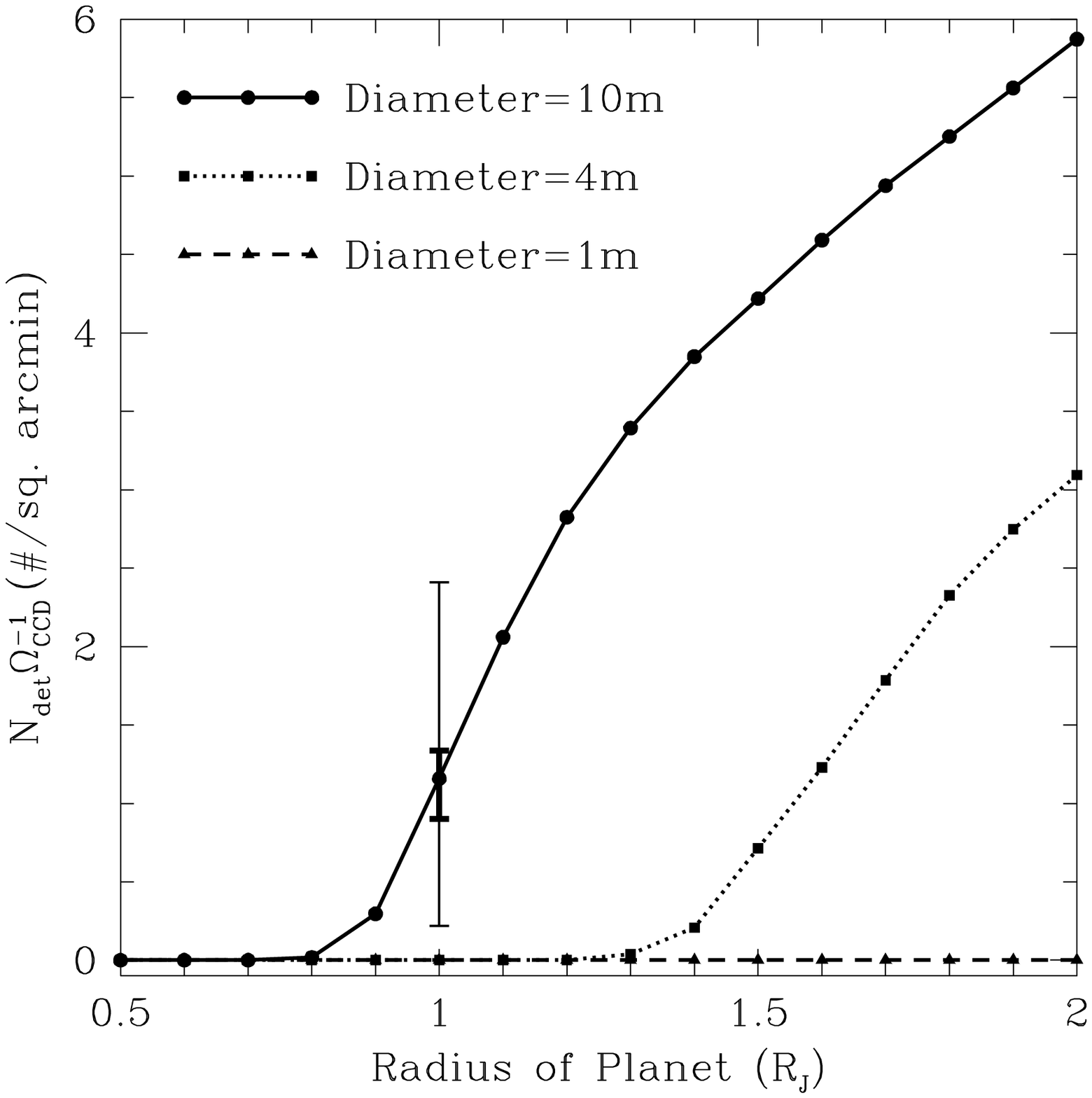}}}}
{\footnotesize {\bf FIG. 4}
The number of planets detected square arcminute as a
function of the radius of the planet for three different telescope
apertures, assuming 10 nights of 8 hours per night.  I have assumed
a seeing of  $\theta=0.75''$, and that 1\% of all stars have planets
distributed uniformly in $\log{a}$ between $a=0.01$ and $0.1$.
The thick errorbar on the point for $D=10$m and $\rp=1\rjup$
corresponds to changing the number of nights by $N=10^{+10}_{-5}$.
The thin errorbar corresponds to changing the seeing by $\theta=[0.75\pm 0.25]''$.}
\bigskip

Figure 4 shows the number of planets detected per unit area
$\ndet \Omega_{CCD}^{-1}$, as a function of $\rp$, for telescope
apertures of $D=10$m, 4m, and 1m, and the fiducial parameters,
$N=10$, $\tw=8$~hours, and $\theta=0.75''$.  For a $5' \times 5'$
field-of-view, a 10m telescope would detect $\sim 100$ planets of
radius $\rp=1.5~\rjup$, and $\sim 30$ planets if $\rp=1.0~\rjup$.  
Most of these planets will be discovered at $a\sim 0.02~\au$ around stars at or slightly
below the turn off:  the number-weighted $I$-magnitude of the sources
for which detections are made is ${\bar I}=19.4$ and the
number-weighted orbital separation is ${\bar a}=0.021$ for
$\rp=1.0\rjup$.  These values approximately constant for $1.0\le \rp/\rjup \le 2.0$.
For small planetary radii, $\rp \la 0.8$, $\ndet < 1$.  Thus if most planets have radii less
than that of Jupiter, it will be quite difficult to detect them around
stars in the Galactic bulge, unless the seeing is excellent, $\theta
\le 0.5''$ (see Fig.\ 4).  For 1m and 4m class telescope, the number of
detected event is almost negligible below $1.5~\rjup$.  Therefore,
such monitoring campaigns are unlikely to detect any planets, unless
there exists a substantial population of companions in the Galactic
bulge with radii $\rp > 1.5\rjup$.

\section{Conclusions}

In this Letter, I have estimated the number of planets that may be detected by
transits in a monitoring campaign toward the Galactic bulge.  An
investment of a relatively modest
amount of telescope resources, 10 clear nights of 8 hours per night on a 10m
telescope at a site with excellent ($0.75''$) median seeing, 
would result in the detection of $\sim 30$ planets of
Jupiter size, if the frequency and distribution of planetary
companions to stars in the Galactic bulge is similar to those of
G-dwarfs in the solar neighborhood.  Most of these planets will be
found at orbital separations of $a\sim 0.02~\au$ around stars slightly 
fainter than the turn-off, i.e.\ evolved
G or early K dwarfs.  Modifications to the observing strategy,
such as decreasing the number of nights to 5 instead of 10, will not result
in substantially fewer detections.  However, if the seeing is
substantially worse than $0.75''$, the number of detections will be
considerably smaller.  Therefore an excellent site is required.
Similar campaigns involving 1m- or 4m-class telescopes are unlikely to
result in any detections toward the bulge. Thus, collaborations
currently monitoring the Galactic bulge for microlensing events are
unlikely to serendipitously detect any transits.

\section*{Acknowledgements}
I would like to thank Andreas Berlind, Andrew Gould, and Paul Martini
for helpful discussions, and Alberto Conti and Penny Sackett for
reading the manuscript.  The original idea for this
paper is due to Penny Sackett. 
This work was supported by an Ohio State University Presidential Fellowship

\end{document}